# A Method to Determine the Tau Neutrino Helicity Using Polarized Taus

M.-T.Dova, L.N.Epele, H.Fanchiotti,
C.A.García Canal, P.E.Lacentre
Departamento de Física
Universidad Nacional de La Plata
1900 La Plata, Argentina

J.D.Swain
Department of Physics
Northeastern University
Boston, Massachusetts 02115, USA

## Abstract

A method is presented to extract the tau neutrino helicity, or equivalently, the chirality parameter $\gamma_{\rm VA}$, independent of any tau polarization which may be present. The method is thus well-suited to measurements using taus produced from the $Z^0$ and is complementary to analyses using tau correlations since it provides the sign of the chirality parameter which is otherwise unavailable without recourse to lower energy experiments where taus are unpolarized. Results of Monte Carlo studies and comments regarding the use of the technique in experiments are also included.

# 1 Introduction

In the Standard Model the $l - W - \nu_l$ vertex is supposed to have the V-A structure for any lepton. This fact has been extensively checked for electrons and muons [1].

Moreover, measurements by the ARGUS Collaboration [2] at energies near the production threshold of $\tau^+ \tau^-$ supported this fact also for the $\tau$ lepton. At higher energies such direct tests have not been made, though the neutral couplings of the $\tau$ lepton to the $Z^0$ have been measured (for a recent summary see [3]). The ALEPH collaboration [4] has performed an analysis suggesting that the $\tau$ charged current is either pure V-A or pure V+A. (Studies of correlations alone do not suffice to determine the sign of $\gamma_{VA}$.)

We analyze here the decay distributions of $e^+e^- \to \tau^+\tau^-$ and $\tau^- \to a_1^- \nu_\tau$ at the $Z^0$ peak with the purpose of obtaining an estimator of the coupling constant of the $\tau - W - \nu_\tau$ vertex. We present an observable for the determination of this constant, which is independent of the value of the $\tau$ polarization.

We use here the single-tau decay modes. Other methods using $\tau^+\tau^-$ spin correlation observables are under study and will be presented in a separate publication [5].

Lepton pairs $\tau^+\tau^-$ are created at LEP from the electron-positron annihilation at energies of the $Z^0$ resonance.

In the Standard Model the neutral current $J^\mu$ has the form:

$$J^\mu = \bar{u}(\tau^+)\gamma^\mu \left(v_\tau - a_\tau \gamma^5\right) u(\tau^-) \qquad (1)$$

where the coupling constants $v_\tau$ and $a_\tau$ are given by:

$$v_\tau = -1 + 4\sin^2\theta_W \qquad (2)$$
$$a_\tau = 1 \qquad (3)$$

The LEP beams are unpolarized but the inequality of the $Z$ couplings to left-handed and right-handed leptons induces a polarization of the taus. The longitudinal polarization of the $\tau^-$ averaged over all the production angles is related to the $\tau^+ - Z^0 - \tau^-$ vertex coupling constants, at the $Z^0$ peak, by:

$$P = -\frac{2 v_\tau a_\tau}{v_\tau^2 + a_\tau^2} \qquad (4)$$

Subsequently, as it is well known, taus decay via weak interactions where parity is not conserved. The $\nu_\tau$ cannot be observed experimentally and the measurable quantities are the energies and momenta of the hadrons or leptons in the final state.

The vertex $\tau - W - \nu_\tau$ is supposed to be $V - A$ in the Standard Model. To take into account a possible deviation from this form we introduce coupling constants $g_V$ and $g_A$ for the vector and axial vector tau currents, namely:

$$J^\mu = \bar{u}(p', s') \gamma^\mu (g_V + g_A \gamma^5) u(p, s) \qquad (5)$$

It is usual to define a quantity analogous to the polarization, which characterizes the handedness of the charged leptonic current. This quantity is called the chirality parameter and is given by:

$$\gamma_{\rm VA} = \frac{2 g_V g_A}{g_V^2 + g_A^2} \qquad (6)$$

In the Standard Model, as it was stated above, $\gamma_{\rm VA} = -1$.

## 2 $\tau^- \to a_1^- \nu_\tau$ Decays

The $a_1$ is a pseudovector resonance decaying into three pions. The decay process is as follows:

$$\tau^- \to a_1^- \nu_\tau, \quad a_1 \to \rho^0 \pi^-, \quad \rho^0 \to \pi^+ \pi^-$$

One can only measure the energies and momenta of the three charged pions in the final state. Only the neutrino escapes the detection so the kinematics of the system is well enough constrained to allow a partial reconstruction of the events, despite the fact that we cannot reconstruct the $\tau$-rest frame. The appropriate frame here is the $a_1$-rest frame where the pions are coplanar as illustrated in Figure 1.

The angular distribution for the decay can be written following reference [6].

$$d\Gamma = N\,[h_1^-\,W_A - (h_1^-\,\cos\theta - h_2\,\sin\theta)\,W_A\,\gamma_{\rm VA}\,P_\tau + h_3\,\cos\beta\,W_E\,P_\tau$$
$$+3\,Q^2\,\cos\psi\,\cos\beta\,W_E\,\gamma_{\rm VA}]\,\frac{(m_\tau^2 - Q^2)^2}{Q^2}$$
$$\frac{dQ^2}{Q^2}\,ds_1\,ds_2\,\frac{d\cos\theta}{2}\,\frac{d\cos\beta}{2} \tag{7}$$

where $q_1$, $q_2$ and $q_3$ are the final pion 4-momenta and $Q = q_1 + q_2 + q_3$. The $\tau$ rest frame decay angle $\theta$ and the angle $\psi$ between the direction of the $\tau$ and the laboratory as seen from the $a_1$ rest frame, can be reconstructed from the energy of the hadronic system,

$$\cos\theta = \frac{2\,x\,m_\tau^2 - m_\tau^2 - Q^2}{(m_\tau^2 - Q^2)\,\sqrt{1 - m_\tau^2/E_{beam}^2}} \quad;\quad \theta \in [0, \pi] \tag{8}$$

$$\cos\psi = \frac{x\,(m_\tau^2 + Q^2) - 2\,Q^2}{(m_\tau^2 - Q^2)\,\sqrt{x^2 - Q^2/E_{beam}^2}} \tag{9}$$

with

$$x = \frac{E_1 + E_2}{E_{beam}} \tag{10}$$

$\beta$ denotes the angle between the normal $n_\perp$ to the three pion plane and the three pions laboratory line of flight. $\cos\beta$ is obtained from the measured pion momenta using the analytic approximation of reference [7]

$$\cos\beta = \frac{8\,Q^2\,\vec{p_1}\cdot(\vec{p_2}\times\vec{p_3})/|\vec{p_1}+\vec{p_2}+\vec{p_3}|}{[-\lambda(\lambda(Q^2,s_1,m_\pi^2),\lambda(Q^2,s_2,m_\pi^2),\lambda(Q^2,s_3,m_\pi^2))]^{\frac{1}{2}}} \tag{11}$$

where $\lambda(x,y,z) = x^2 + y^2 + z^2 - 2xy - 2yz - 2zx$.

$$h_1^\pm = m_\tau^2 \pm 2\,Q^2 - (m_\tau^2 \mp Q^2)\,\frac{3\cos^2\psi - 1}{2}\,\frac{3\cos^2\beta - 1}{2} \tag{12}$$

$$h_2 = 3\,m_\tau\,\sqrt{Q^2}\,\frac{\sin 2\psi}{2}\,\frac{3\cos^2\beta - 1}{2} \tag{13}$$

$$h_3 = -3\,Q^2\,(\cos\theta\,\cos\psi + \frac{m_\tau}{\sqrt{Q^2}}\,\sin\theta\,\sin\psi) \tag{14}$$

$$N = \frac{G^2}{8\,m_\tau^3}\,(g_V^2 + g_A^2)\,\cos^2\theta_C\,\frac{1}{64\,(2\,\pi)^5} \tag{15}$$

and

$$W_A = |F_1|^2 \left[(s_2 - 4 m_\pi^2) + \frac{(s_3 - s_1)^2}{4 Q^2}\right]$$
$$+ |F_2|^2 \left[(s_1 - 4 m_\pi^2) + \frac{(s_3 - s_2)^2}{4 Q^2}\right]$$
$$+ \left[(Q^2 - 2 s_3 - m_\pi^2) + \frac{(s_3 - s_1)(s_3 - s_2)}{2 Q^2}\right] Re(F_1 F_2^*) \quad (16)$$

$$W_E = 3 \left[\frac{s_1 s_2 s_3 - m_\pi^2 (Q^2 - m_\pi^2)^2}{q^2}\right]^{1/2} Im(F_1 F_2^*) \quad (17)$$

The Dalitz variables $s_1$ and $s_2$ are defined by

$$s_i = (q_j + q_+)^2 \; ; \; i \neq j = 1, 2 \quad (18)$$

where $q_+$ is the momentum of the positive pion.

The model dependence is contained in the functions $W_A$ and $W_E$. The discussion presented by J.H.Kuhn and E.Mirkes [6] is based on a hadronic current of the form:

$$J^\mu = F_1(s_1, s_2, Q^2) \; V_1^\mu + F_2(s_1, s_2, Q^2) \; V_2^\mu \quad (19)$$

with

$$V_1^\mu = q_1^\mu - q_3^\mu - Q^\mu \frac{Q(q_1 - q_3)}{Q^2} \quad (20)$$

$$V_2^\mu = q_2^\mu - q_3^\mu - Q^\mu \frac{Q(q_2 - q_3)}{Q^2} \quad (21)$$

$F_1 = F(s_1, s_2, Q^2) \; ; \; F_2 = F(s_2, s_1, Q^2)$

and a choice for F as follows:

$F(s_1, s_2, Q^2) = -\frac{2\sqrt{2}i}{3 f_\pi} BW_a(Q^2) B_\rho(s_2),$

where $BW_a$ and $B_\rho$ denote Breit - Wigner resonances.

This model for the current has previously been worked out by J.H.Kuhn and F.Wagner [8] and is implemented in the KORALZ event generator [9] widely used to simulate $\tau$ production and decays.

Given that the two negative pions are not distinguishable, there are two possible ways to form the $\rho$-meson. The interference between them is contained in the function $W_E$ through the imaginary part of the structure functions $F_1$ and $F_2$. Notice that the only term in the angular distribution that contain $\gamma_{\mathrm{VA}}$ without the presence of $P_\tau$ is proportional to $W_E$. This is the interference that makes the $\tau \to a_1 \nu_\tau$ the unique hadronic channel from which we can disentangle the dependence on the chirality parameter.

## 3  Determination of the Chirality Parameter

A method to obtain an estimator of the chirality parameter, which is model dependent though, consists in taking appropriate moments using the distribution function given in equation (7). To go further along this new method, let us introduce the following notation

$$d\Gamma = \Omega(Q^2, s_1, s_2, \cos\theta, \cos\beta) \, d^5x \qquad (22)$$

$$d^5x = dQ^2 \, ds_1 \, ds_2 \, \frac{d\cos\theta}{2} \, \frac{d\cos\beta}{2} \qquad (23)$$

Then for any quantity $m$, we define a moment $\langle m \rangle$ by

$$\langle m \rangle = \frac{\int m \, \Omega(Q^2, s_1, s_2, \cos\theta, \cos\beta) \, d^5x}{\int \Omega(Q^2, s_1, s_2, \cos\theta, \cos\beta) \, d^5x} \qquad (24)$$

The important observation is that it is possible to eliminate the dependence on the $\tau$ polarization by taking the moment of the quantity

$$\mathcal{M} = \frac{\cos\beta \, \cos\theta \, sgn(s_1 - s_2)}{\cos\theta \, \cos\psi + \frac{m_\tau}{\sqrt{Q^2}} \sin\theta \, \sin\psi} \qquad (25)$$

The function $sgn$ of $(s_1 - s_2)$ is introduced in order to take into account the ambiguity in the direction of the normal to the decay plane, due to the Bose symmetry of the two negative pions. Then one has

$$\int \mathcal{M} \, \Omega \, ds_1 \, ds_2 \qquad \frac{d\cos\theta}{2} \, \frac{d\cos\beta}{2}$$

$$= N \gamma_{\text{VA}} \frac{(m_\tau^2 - Q^2)^2}{Q^2} \int \frac{\cos\theta \cos\psi}{\cos\theta \cos\psi + \frac{m_\tau}{\sqrt{Q^2}} \sin\theta \sin\psi}$$

$$W_E \, sgn(s_1 - s_2) \, ds_1 \, ds_2 \, \frac{d\cos\theta}{2} \qquad (26)$$

while the normalization constant $N$ is determined by the condition

$$1 = \int \Omega \, ds_1 \, ds_2 \, \frac{d\cos\theta}{2} \, \frac{d\cos\beta}{2}$$

$$= N \frac{(m_\tau^2 - Q^2)^2}{Q^4} (m_\tau^2 + 2Q^2) \int W_A \, ds_1 \, ds_2 \qquad (27)$$

Finally one can write

$$\langle \mathcal{M} \rangle = -\gamma_{\text{VA}} \, A_{LR}(Q^2) \, T(Q^2) \qquad (28)$$

where we have introduced the functions

$$A_{LR} = -\frac{Q^2}{(m_\tau^2 + 2Q^2)} \frac{\int W_E \, sgn(s_1 - s_2) \, ds_1 \, ds_2}{\int W_A \, ds_1 \, ds_2} \qquad (29)$$

$$T(Q^2) = -\frac{1}{(m_\tau^2 - Q^2)} \{[Q^2 + m_\tau^2 \, [1 + \frac{3 m_\tau^2 + Q^2}{3 K(Q^2)} \log \frac{3 m_\tau^2 - Q^2 - K(Q^2)}{3 m_\tau^2 - Q^2 + K(Q^2)}$$

$$+ \log \frac{m_\tau}{\sqrt{Q^2}}]\} \qquad (30)$$

with

$$K(Q^2) = [(9 m_\tau^2 - Q^2)(m_\tau^2 - Q^2)]^{1/2} \qquad (31)$$

## 4 Monte Carlo Studies

We have performed a Monte Carlo study using the Koralz [9] program to generate samples of 200,000 events with $a_1$ decays assuming pure V-A and pure V+A charged current couplings, as well as 200,000 events with non-standard values of $\gamma_{\text{VA}}$ to represent a hypothetical data sample with $g_V = 0.6$ and $g_V^2 + g_A^2$ unchanged from its standard model value, giving $\gamma_{\text{VA}} = -0.768$. The calculated values of the moments and their errors are shown in Figure 2 for each of the three data samples. A $\chi^2$ fit for the best linear combination of V-A and V+A samples to match the $\gamma_{\text{VA}} = -0.768$ sample gave a statistical error of 0.049, which includes errors due to the finite Monte Carlo V-A and V+A samples as well as those due to the finite number of events with nonstandard couplings. Monte Carlo studies using samples of fully right or left-handedly polarized taus give consistent answers, verifying that the method of this papers gives a method for the determination of the tau neutrino chirality parameter which is independent of the tau polarization.

The errors are, admittedly, large when scaled to realistic numbers of events at LEP, but two points are worth bearing in mind : 1) studies of correlated tau decays at LEP can give quite accurate determinations of the absolute value of the tau neutrino chirality parameter, but with absolutely no information about its sign. The information from the correlated decays and the method in this paper are statistically independent, and thus the likelihood distributions for the tau neutrino chirality parameter can be multiplied. The final likelihood will be quite sensitive to both the sign and magnitude of $\gamma_{\rm VA}$. 2) Future accelerators (and perhaps higher luminosity options for LEP) may provide large enough data samples so that the intrinsic interest of this method (without recourse to studies of correlated tau decays) will be greater. In particular, the method can be used to study singly-produced tau decays from hadron machines.

Clearly in taking moments to remove the tau polarization dependence we have assumed a perfect detector. In any real experiment, some polarization dependence in the calculated moments is bound to appear and must be studied. In addition, we are studying the possibility of using a related method fixing the product of the tau polarization and tau neutrino chirality parameters [10].

One final comment of interest for experiments is that a sample of $\tau^-$ decays with a V+A charged current interaction can be obtained simply by using $\tau^+$ decays with a V-A interaction and simply reversing the signs of the charges of all particles.

## 5  Summary

In summary, we have found an observable for the determination of $\gamma_{\rm VA}$ which is independent of the $\tau$ polarization. The function $A_{LR}(Q^2)$ is the parity vi-olating asymmetry measured by the ARGUS collaboration to determine the chirality at low energies where $P_\tau = 0$. The method has been checked with simulated events generated by Monte Carlo and the sensitivity estimated.

This method can also be applied to the process in which $a_1^- \to \pi^0\, \pi^0\, \pi^-$. To this end one has only to change $q_+ \to q_-$ in equation (18) and $q_j$ ; $j = 1, 2$ are now the neutral pion momenta. Clearly, in this case one is dealing with a negative $\rho$-meson.

# 6 Acknowledgements

This work was supported in part by Consejo Nacional de Investigaciones Científicas y Técnicas (CONICET), the National Science Foundation, and the World Laboratory Project. M.-T.D. would like to thank CERN for its hospitality while part of this work was being done. J.D.S. would like to thank the Universidad Nacional de La Plata for its kind hospitality when this work was being finished.

# References

[1] K. Mursula, M. Roos, F. Scheck, Nucl. Phys. **B219** (1983) 321. W. Fetscher, H.J. Gerber, K.F. Johnson, Phys. Lett. **B173** (1986) 102.

[2] ARGUS Collaboration, H. Albrecht et al., Phys. Lett. **B250** (1990) 164. ARGUS Collaboration, H. Albrecht et al., DESY preprint 94-120 (1994).

[3] J. Swain, "Weak Couplings of the $\tau$ Lepton : Results from LEP", invited summary talk at the American Physical Society, DPF Meeting, Albuquerque, New Mexico, August 1-6, 1994 (to appear in the proceedings). Available as Northeastern University preprint NUB-3100.

[4] D. Buskulic et al., Phys. Lett. **B321** (1994) 168.

[5] M.-T. Dova et al., "Correlated Angular Distributions in $\tau^+\tau^-$ Semileptonic Decays", La Plata preprint, submitted for publication.

[6] J.H. Kühn and E. Mirkes, Phys. Lett. **B286** (1992) 381.

[7] A. Rougé, Workshop on Tau Lepton Physics Orsay 1990, Editions Frontiers (1991) 213.
A. Rougé, Z. Phys. **C48** (1990) 75.

[8] J.H.Kühn and F.Wagner, Nucl. Phys. **B236** (1984) 16.

[9] S. Jadach and Z. Was, *Comput. Phys. Commun.* **35** (1985);
R. Kleiss, "Z Physics at LEP", CERN-8908 (1989), Vol. III, p. 1.

[10] M.-T. Dova et al., in preparation.

**Figure Captions**

**Figure 1:** Kinematics for the $a_1$ decay.

**Figure 2:** Moments vs. $Q^2$ for various values of $\gamma_{\text{VA}}$.

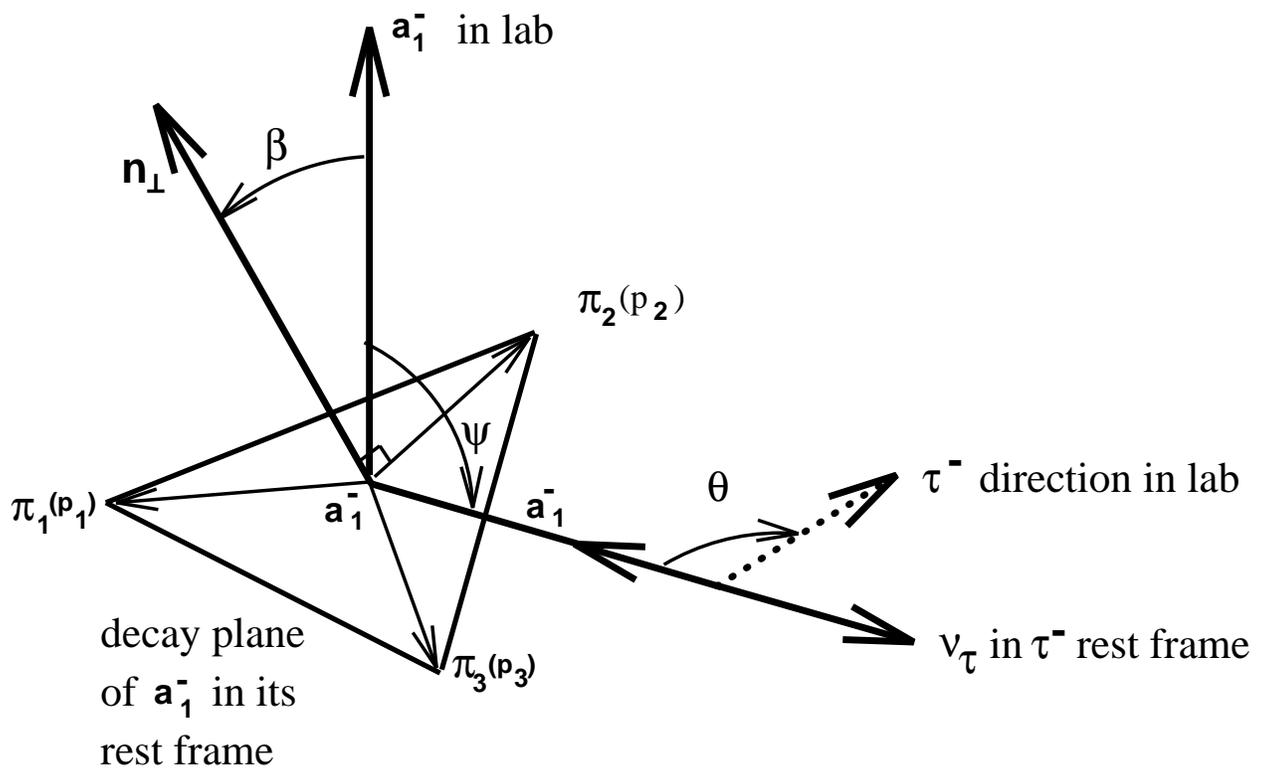

**Figure 1.**

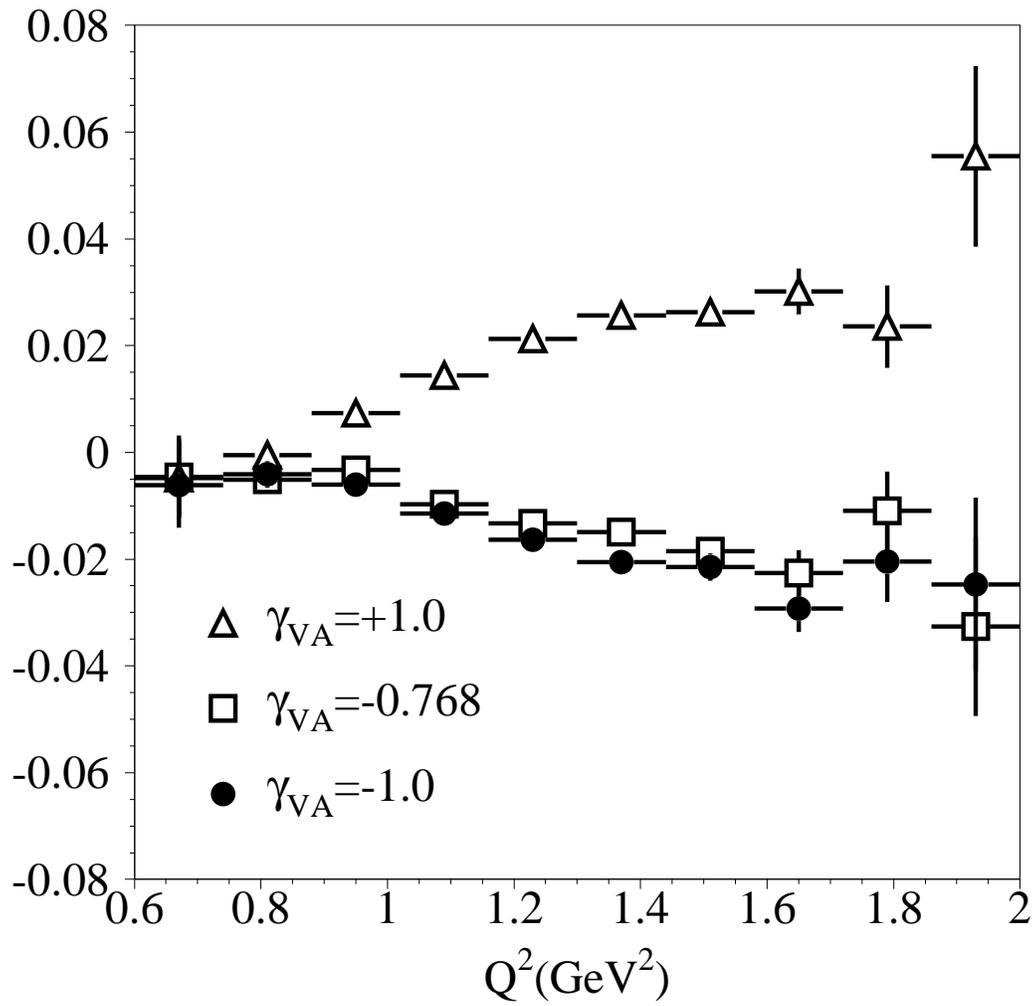

**Figure 2.**